\documentclass[%
 reprint,
superscriptaddress,
 amsmath,amssymb,
 aps,
prl,
]{revtex4-1}

\usepackage{graphicx}% Include figure files
\usepackage{epstopdf}
\usepackage{dcolumn}% Align table columns on decimal point
\usepackage{bm}% bold math
\usepackage{gensymb}
\usepackage{upgreek}
\usepackage{hyperref}
\usepackage{graphicx}% Include figure files
\usepackage{epstopdf}
\usepackage{dcolumn}% Align table columns on decimal point
\usepackage{bm}% bold math
\usepackage{gensymb}
\usepackage{upgreek}
\usepackage{hyperref}
\usepackage{lineno}
\usepackage{xr}
\makeatletter
\newcommand*{\addFileDependency}[1]{
  \typeout{(#1)}
  \@addtofilelist{#1}
  \IfFileExists{#1}{}{\typeout{No file #1.}}
}
\makeatother

\newcommand*{\myexternaldocument}[1]{
    \externaldocument{#1}
    \addFileDependency{#1.tex}
    \addFileDependency{#1.aux}
}
\myexternaldocument{FeP-SI}

\begin{document}

\preprint{APS/123-QED}

\title{Topologically-Driven Linear Magnetoresistance in Helimagnetic FeP}

\author{D.J. Campbell}
\thanks{daniel.campbell@lncmi.cnrs.fr, Present address: LNCMI, CNRS, EMFL, Univ. Grenoble Alpes, INSA Toulouse, Univ. Toulouse Paul Sabatier, 38000 Grenoble, France}
\affiliation{Maryland Quantum Materials Center, Department of Physics, University of Maryland, College Park, Maryland 20742, USA}
\author{J. Collini}
\affiliation{Maryland Quantum Materials Center, Department of Physics, University of Maryland, College Park, Maryland 20742, USA}
\affiliation{NIST Center for Neutron Research, NIST, Gaithersburg, Maryland 20899, USA}
\author{J. S\l{}awi\'{n}ska}
\thanks{Present address: Zernike Institute for Advanced Materials, University of Groningen, Nijenborgh 4, 9747AG, Netherlands}
\affiliation{Department of Physics, University of North Texas, Denton, TX 76203, USA}
\author{C. Autieri}
\affiliation{International Research Centre Magtop, Institute of Physics, Polish Academy of Sciences, Aleja Lotnik\'{o}w 32/46, PL-02668 Warsaw, Poland}
\affiliation{Consiglio Nazionale delle Ricerche CNR-SPIN, UOS Salerno, I-84084 Fisciano (Salerno), Italy}
\author{L. Wang}
\affiliation{Maryland Quantum Materials Center, Department of Physics, University of Maryland, College Park, Maryland 20742, USA}
\author{K. Wang}
\affiliation{Maryland Quantum Materials Center, Department of Physics, University of Maryland, College Park, Maryland 20742, USA}
\author{B. Wilfong}
\affiliation{Maryland Quantum Materials Center, Department of Physics, University of Maryland, College Park, Maryland 20742, USA}
\affiliation{Department of Chemistry, University of Maryland, College Park, Maryland 20742, USA}
\author{Y.S. Eo}
\affiliation{Maryland Quantum Materials Center, Department of Physics, University of Maryland, College Park, Maryland 20742, USA}
\author{P. Neves}
\affiliation{Maryland Quantum Materials Center, Department of Physics, University of Maryland, College Park, Maryland 20742, USA}
\affiliation{NIST Center for Neutron Research, NIST, Gaithersburg, Maryland 20899, USA}
\author{D. Graf}
\affiliation{National High Magnetic Field Laboratory, 1800 East Paul Dirac Drive, Tallahassee, Florida 32310, USA}
\author{E.E. Rodriguez}
\affiliation{Maryland Quantum Materials Center, Department of Physics, University of Maryland, College Park, Maryland 20742, USA}
\affiliation{Department of Chemistry, University of Maryland, College Park, Maryland 20742, USA}
\author{N.P. Butch}
\affiliation{Maryland Quantum Materials Center, Department of Physics, University of Maryland, College Park, Maryland 20742, USA}
\affiliation{NIST Center for Neutron Research, NIST, Gaithersburg, Maryland 20899, USA}
\author{M. Buongiorno Nardelli}
\affiliation{Department of Physics, University of North Texas, Denton, TX 76203, USA}
\author{J. Paglione}
\thanks{paglione@umd.edu}
\affiliation{Maryland Quantum Materials Center, Department of Physics, University of Maryland, College Park, Maryland 20742, USA}
\affiliation{Canadian Institute for Advanced Research, Toronto, Ontario M5G 1Z8, Canada}

\date{\today}

\begin{abstract}
The  helimagnet FeP is part of a family of binary pnictide materials with the MnP-type structure which share a nonsymmorphic crystal symmetry that preserves generic band structure characteristics through changes in elemental composition. It shows many similarities, including in its magnetic order, to isostructural CrAs and MnP, two compounds that are driven to superconductivity under applied pressure. Here we present a series of high magnetic field experiments on high quality single crystals of FeP, showing that the resistance not only increases without saturation by up to several hundred times its zero field value by 35~T, but that it also exhibits an anomalously linear field dependence over the entire field range when the field is aligned precisely along the crystallographic \textit{c}-axis. A close comparison of quantum oscillation frequencies to electronic structure calculations links this orientation to a semi-Dirac point in the band structure which disperses linearly in a single direction in the plane perpendicular to field, a symmetry-protected feature of this entire material family. We show that the two striking features of MR\textemdash{}large amplitude and linear field dependence\textemdash{}arise separately in this system, with the latter likely due to a combination of ordered magnetism and topological band structure.
\end{abstract}

\maketitle

%%%%%%%%%%%%%%%%%%%%%%%%%%%%%%%%%%%%%%%%%%%%
\section{\label{sec:Intro}Introduction}

The orthorhombic MnP-type (or B31) family of materials has been under study for several decades \cite{SelteFeAs,KallelHelimagnetism,WesterstrandhFeP,TakaseMnPHiField} but its diverse range of properties has recently been the subject of renewed attention. Aside from a peripheral connection to iron-based high-temperature superconductivity \cite{PaglioneFeSCs}, novel magnetism \cite{ParkerFeAs,RodriguezFeAsSDW,CampbellFeAs}, quantum criticality \cite{ChengCrAsMnP,MatsudaCrAsQC}, metal-insulator transitions \cite{HiraiRuRhPn}, and indications of non-trivial electronic topology \cite{LvTaAs,ShekharNbP,NiuCrAs} have all been reported in a series of binary transition metal-pnictides. Two members of the B31 family, CrAs and MnP itself, have also been shown to superconduct upon suppression of helimagnetic order under applied pressure \cite{KotegawaCrAs,WuCrAs,ChengMnP,ChengCrAsMnP}, suggesting a novel interplay of ground states. Furthermore, a linear magnetoresistance (MR) was observed in CrAs near the magnetic quantum critical point and was attributed to the presence of a ``semi-Dirac'' point in the band structure (one which disperses linearly along a single direction in momentum space) \cite{NiuCrAs}. It has recently been shown that the nonsymmorphic $Pnma$ structure of the MnP family preserves many specific band structure features, including the semi-Dirac point, across different members \cite{CuonoMnPType}. CrAs, MnP, and paramagnetic WP (ambient pressure $T_c = 0.8~$K) \cite{LiuWP} are all predicted to be unconventional topological superconductors as a result, suggesting a possible connection between helimagnetism, superconductivity, and non-trivial topological features that deserves further attention.

Magnetoresistance (MR) has been key to revealing topological properties in many other materials. Semimetals such as WTe$_2$ \cite{AliWTe2}, Cd$_3$As$_2$ \cite{FengCd3As2}, or NbP \cite{ShekharNbP} have shown extremely large ($\mathcal{O}(10^5)$ enhancement over $\rho{}_{0\textrm{T}}$), nonsaturating increases of the resistivity in field. This has been attributed to high mobility, massless carriers that result from the linear dispersion. However, for these materials semiclassical explanations could also be valid for the observed phenomena \cite{FauqueAntimony}. For this reason it is important to find materials with topological band structure elements but with other characterstics that differentiate them from the typical Weyl/Dirac semimetal. The very good metallic behavior and ordered magnetism found in the MnP-type family provide such an opportunity.

Here we present electrical resistance measurements of high quality single crystals of FeP, a B31 family member isostructural to CrAs and MnP, which orders magnetically below $T_N$~=~120~K \cite{KallelHelimagnetism}, in a state which was shown in closely related FeAs to feature a non-collinear spin-density wave order \cite{RodriguezFeAsSDW}. Following up on basic transport and physical property measurements \cite{WesterstrandhFeP,SelteFeP}, we focus on transport and fermiology under high magnetic fields. We observe large, nonsaturating magnetoresistance reaching values of several hundred times, which we attribute to a shift in carrier mobility below approximately 50~K that causes increased compensation of the Fermi surface (FS), an explanation that likely extends to other B31 compounds. Studying the field-angle dependence, we observe features of a complex Fermi surface but also a singular linear MR when field is directed precisely along the crystallographic \textit{c}-axis. A careful comparison of quantum oscillations data with calculated band structure directly confirms the location of the semi-Dirac point in this system, and more importantly, its role in the anomalous linear MR that would only occur in a certain field orientation. Seizing on new theoretical work \cite{XiaoLMR}, we propose a link between linear MR, topological band structure, and the magnetically ordered state, which combine to unique effect in FeP.

%%%%%%%%%%%%%%%%%%%%%%%%%%%%%%%%%%%%%%%%%%%%
\section{Results}

\subsection{Magnetotransport}

\begin{figure}
    \centering
    \includegraphics[width=0.45\textwidth]{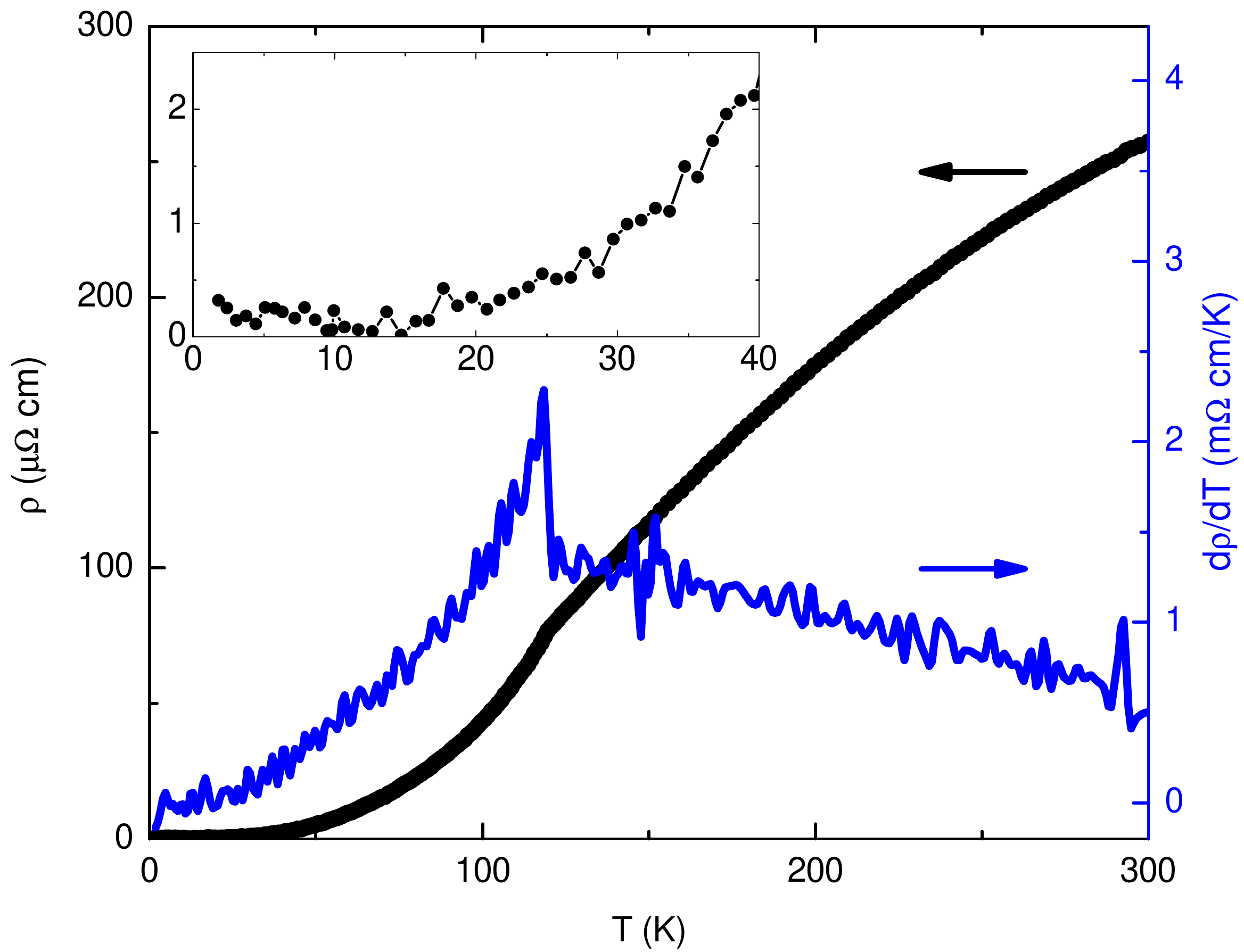}
    \caption{Electrical resistivity (black, left axis) and its derivative (blue, right axis) as a function of temperature for an FeP crystal with a residual resistance ratio of 1000 and B = 0~T. The kink in the resistivity and jump in the derivative at 120~K correspond to the N\'{e}el temperature. The inset zooms in on the low temperature resistivity, highlighting the plateau below 20~K and value of residual resistivity of about 0.2~$\mu{}\Omega{}$~cm.}
    \label{fig:FePRhoVST}
\end{figure}

The electrical resistivity (Fig.~\ref{fig:FePRhoVST}) and magnetic susceptibility \cite{WesterstrandhFeP} of FeP are very similar to that of FeAs \cite{SegawaFeAs}, with S-shaped curvature and distinct kink at $T_N$ (which is 70~K in the arsenide), especially noticeable in the derivative, suggesting that itinerant spin-density wave magnetism is very similar in both compounds. However, even compared to the highest quality crystals of FeAs \cite{CampbellFeAs}, the residual resistivity of FeP is extremely small (as low as 0.2~$\mu{}\Omega{}$~cm), with a residual resistivity ratio (RRR = $\rho{}_{\textrm{300~K}}/\rho{}_{\textrm{1.8~K}}$) of up to 1500, much larger than that of CrP and CrAs \cite{WuCrAs,NiuCrP,NigroCrAsMR}, and rivaled only by MnP  \cite{TakaseMnPRRR}. The trend of higher RRR in phosphides appears generic to this family, as it is also observed in CoAs \cite{CampbellCoAs} and CoP [SI, Fig.~S\ref{fig:CoPTransport}]. This makes FeP an ideal candidate for high fidelity measurements of magnetotransport and quantum oscillations, in particular at the high fields available with resistive magnets. 

\begin{figure}
    \centering
    \includegraphics[width=0.45\textwidth]{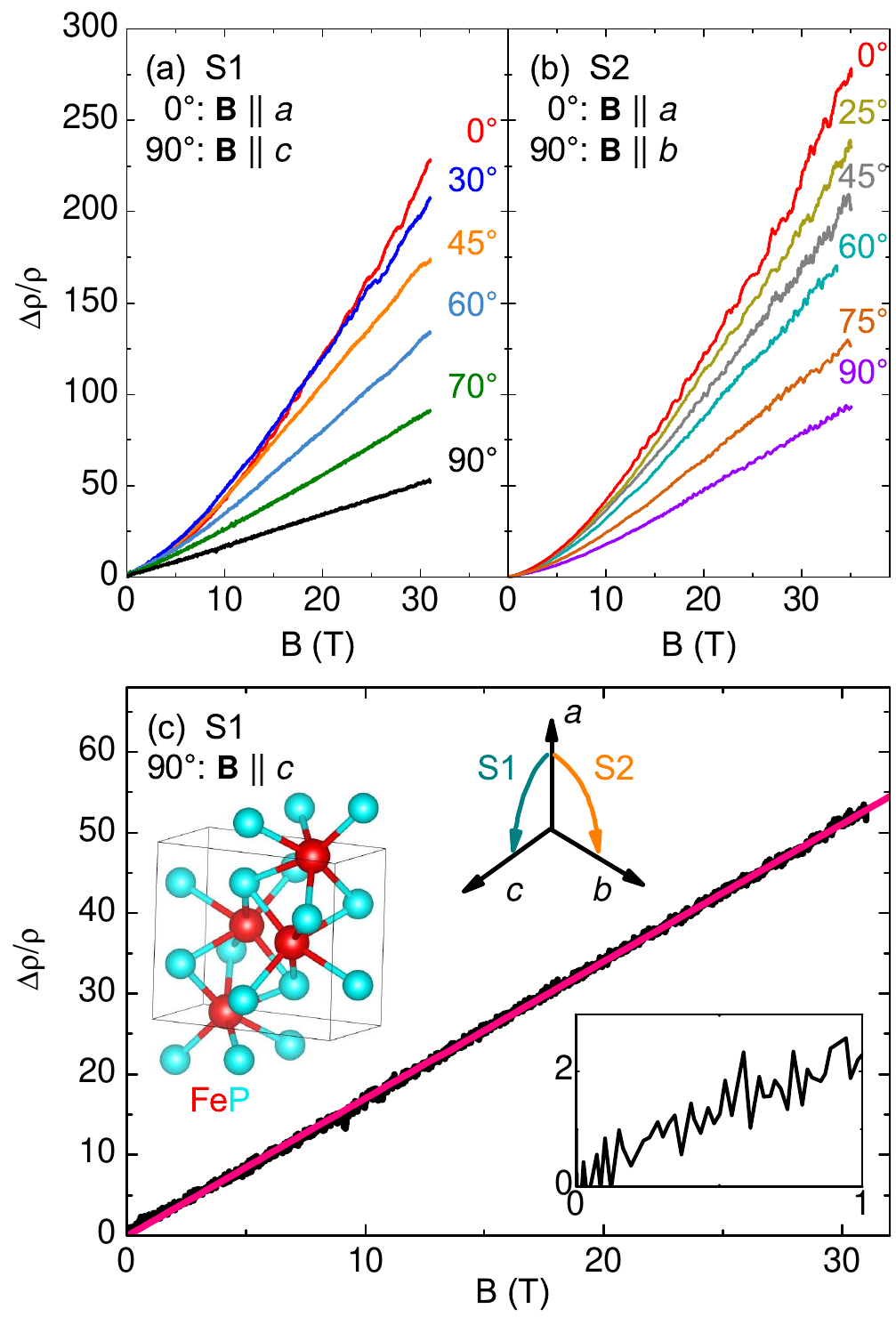}
    \caption{Transverse magnetoresistance $\Delta{}\rho{}/\rho{}$, defined as $\frac{\rho{}(B) - \rho{}(0~\textrm{T})}{\rho{}(0~\textrm{T})}$, measured up to high magnetic fields oriented at several angles for FeP crystals (a) S1 (RRR~$\approx{}$~1300) and (b) S2 (RRR~$\approx{}$~1400), whose orientations are noted. Note that S1 was actually aligned with two different reflections of the [101] family, but the angle has been shifted to ease readability. The temperature was held constant at 400~$\pm~50$~mK for all measurements. MR is larger for S2 at the common \textbf{B}~$\parallel$~\textit{a}-axis orientation, which can be linked to its higher RRR value. (c) The \textbf{B}~$\parallel$~\textit{c}-axis data for S1, with a linear fit (pink) made over the entire field range.  The graphics above the curve show an image of the unit cell highlighting the octahedral Fe-P coordination. The axes indicate the alignment of the crystal picture and how field was rotated for S1 and S2 (current was always perpendicular to field in either case). The inset below the curve is a closeup of the data below 1~T.}
    \label{fig:FePHiFieldMR}
\end{figure}

We focus on two single-crystal samples, S1 and S2, for magnetotransport measurements with fields rotated through different crystallographic orientations as shown in Fig.~\ref{fig:FePHiFieldMR}. Both S1 and S2 exhibit very large and nonsaturating MR at all angles, as well as multi-frequency quantum oscillations (QOs). For both, the largest MR is observed when \textbf{B}~$\parallel$~[100]. It is slightly larger for S2, likely due to its higher RRR \cite{AliWTe2,ChenMgB2RRRMR, PippardMRinMetals}. Most angles show generally similar behavior, with a superlinear field dependence that becomes more linear at high applied field. Power law fits to MR data below 15~T have a maximum $n = 1.5$ for the angles exhibiting the largest MR, which is nearly 300 times the zero field resistivity. This increase is more than two orders of magnitude larger than that of in FeAs in high field \cite{CampbellFeAs}. Given the close structural and magnetic parallels between the two materials, it seems that the substitution of P for As and related significant decrease in residual resistivity produce a bigger in-field response. Indeed, the B31 materials with the largest MR are all phosphides. MnP \cite{TakaseMnPHiField} has only been measured up to 8~T, but at some angles actually shows a larger increase than FeP up to that field. CrP \cite{NiuCrP} has been measured up to 58~T with an MR of about 350. All of these have a significant angular variation of MR. There are other examples of large MR in transition metal pnictide binaries: the four (Nb/Ta)(P/As) combinations, which form in a cubic structure, can reach values exceeding 1000 by 10~T. A comparison of the maximum reported MR for a variety of transition metal-pnictogen binaries is given in Table~I of the SI.

\begin{figure}
    \centering
    \includegraphics[width=0.45\textwidth]{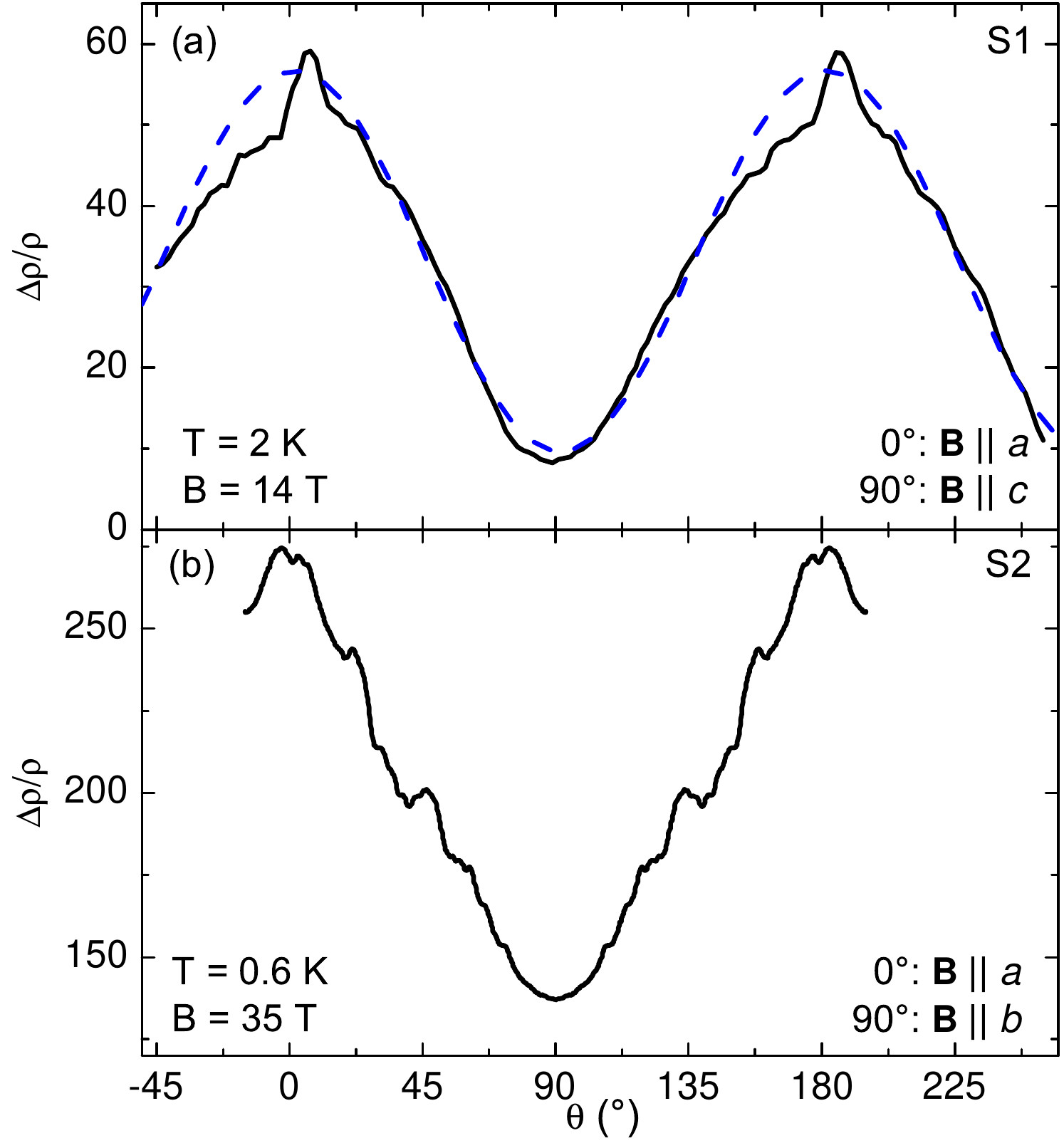}
    \caption{Angle-dependent magnetoresistance for FeP samples (a) S1 and (b) S2 from Fig.~\ref{fig:FePHiFieldMR} at magnetic fields of 14~T and 35~T, respectively. The dashed line in (a) is a cosine fit to the data with only the period fixed, and the spike at maximum MR excluded. Data in (b) were originally taken only from -15$\degree{}$ to 110$\degree{}$, and so were mirrored around 90$\degree{}$.}
    \label{fig:FePRotator}
\end{figure}

The most striking aspect of the FeP angular MR data are found when the magnetic field is aligned along the crystallographic $c$-axis. As shown in Fig.~\ref{fig:FePHiFieldMR}(c), the MR data are linear from zero field [Fig.~\ref{fig:FePHiFieldMR}(c) inset] up to the highest measured field of 35~T. The low-field behavior was also verified in the same sample after measurements at NHMFL, confirming the linearity when sweeping the field through zero [see supplementary Fig.~S\ref{fig:Kohler}]. This orientation also has the lowest MR of any of the angles measured in either rotation plane. While MR tends toward linear behavior at high fields for all angles, there is no smooth decrease in crossover field, as none of the other curves are truly linear even below 10~T. Therefore the $c$-axis MR linearity must be closely linked to a particular property of the Fermi surface at that orientation.

The anisotropy is seen more clearly in a measurement of MR upon field rotation at constant fields, shown in Fig.~\ref{fig:FePRotator}, which also reveals intricate features similar to those observed in MnP \cite{TakaseMnPHiField}. Both these features and the underlying $\pi{}$-periodic dependence indicate an anisotropic band structure with complex Fermi pocket shapes. Interestingly, as shown in Fig.~\ref{fig:FePRotator}(a) the maximum in MR actually occurs at an angle 5$\degree$ away from the \textit{a}-axis, where there is a narrow spike. This is not due to misalignment of the rotator, because the resistance minimum lies near 90$\degree$ and is smoother and more symmetric. In fitting a cosine to the data and leaving all parameters except the $\pi{}$ periodicity free, the maximum is still within 2$\degree$ of the expected location. 

\begin{figure}
    \centering
    \includegraphics[width=0.45\textwidth]{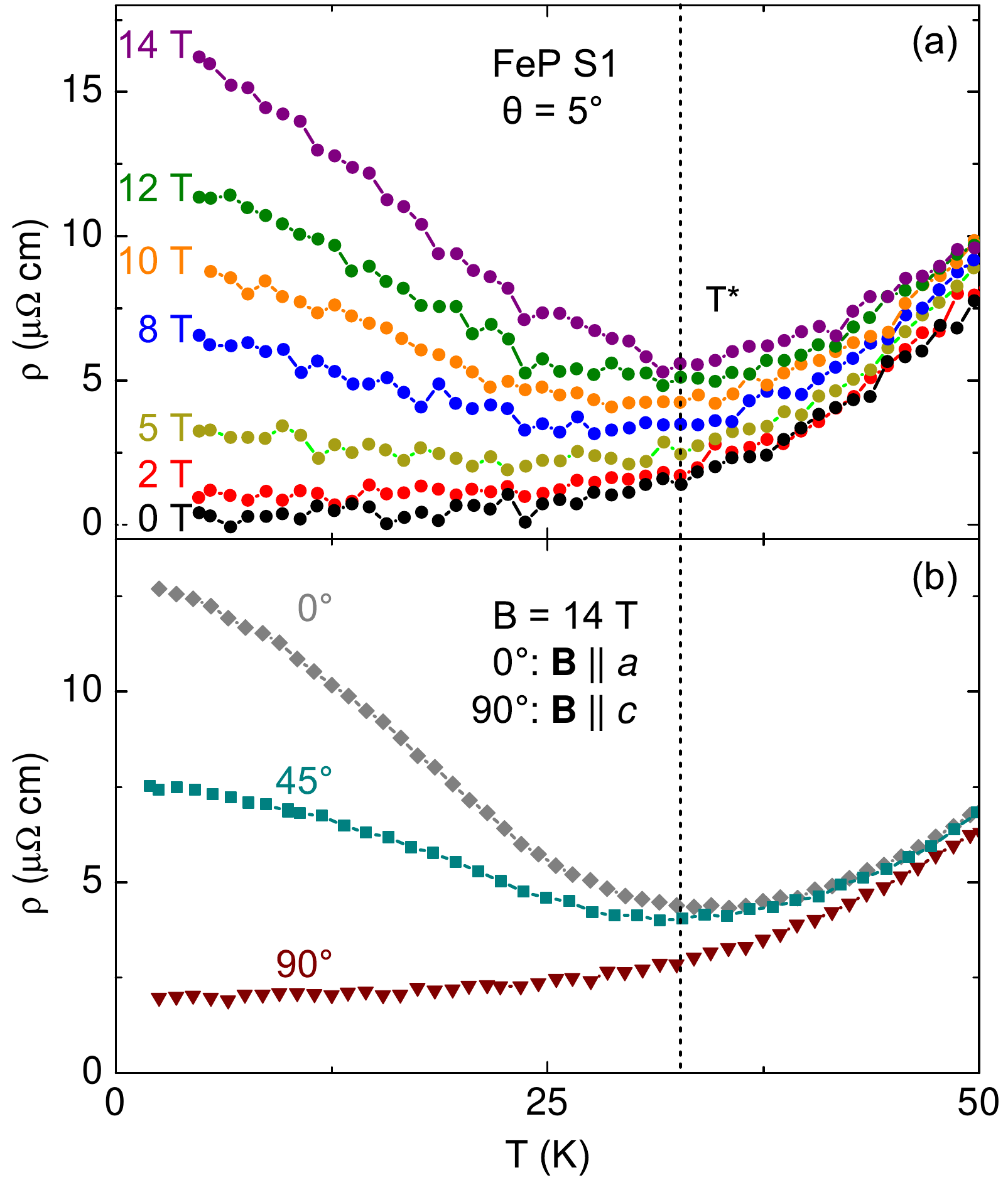}
    \caption{Temperature dependence of the resistivity of sample S1 for different orientations and magnetic field strengths. (a) Temperature dependence at the maximal magnetoresistive angle for field in the \textit{a}-\textit{c} plane ($\theta = 5$\degree{}) from Fig.~\ref{fig:FePRotator}(a), in zero and various applied fields. (b) Temperature-dependent MR of the same sample at three more angles and 14~T; for all three, at least the beginning of a plateau is apparent at low temperature, in contrast to the $\theta = 5$\degree{}, 14~T data in panel (a).}
    \label{fig:FePTDependence}
\end{figure}

To further investigate the detailed MR behavior, temperature sweeps at constant fields up to 14~T were made for specific angles, showing similar behavior to that seen in CrP and several extreme MR rare-earth pnictide binaries, with a ``turn on'' temperature, \textit{T*}, below which magnetoresistance increases significantly \cite{NiuCrP, AliWTe2}. Figure~\ref{fig:FePTDependence}(a) presents data for the angle with maximum MR in Fig.~\ref{fig:FePRotator}(a), 5$\degree$ away from \textbf{B}~$\parallel$~\textit{a}. Here we define \textit{T*} as the temperature of the resistance minimum in the 14~T sweep, roughly 35~K. As demonstrated in Fig.~\ref{fig:FePTDependence}(b), below this temperature the angular dependence of the MR becomes noticeable, similar to that of previously mentioned materials.

As can be seen in both Fig.~\ref{fig:FePRhoVST} and Fig.~\ref{fig:FePTDependence}(a), the zero field resistance changes little below about 20~K. Our observation of a constant  linear increase in MR down to 2~K means that in this temperature range the MR at constant \textit{B} is no longer solely a function of $\rho{}_0$, in violation of Kohler's rule \cite{WangWTe2Upturn,JoPdSn4}. In other words, a different scattering mechanism has emerged below \textit{T*}, or different parts of the FS are contributing to scattering. This can be compared to the data from other angles [Fig.~\ref{fig:FePTDependence}(b)], all of which have a minimum at a similar \textit{T*} at 14~T and at least sublinear behavior at lowest temperature, indicating an approach to saturation, mimicking zero-field behavior. This includes the data set taken at 0$\degree$, a small shift from that in Fig.~\ref{fig:FePTDependence}(a). Thus the Kohler's rule violation does not occur at all angles (See Fig.~S\ref{fig:Kohler} of the Supplemental Information, which shows Kohler's rule obeyed for \textbf{B}~$\parallel$~\textit{c}-axis).

Field sweeps in a Hall geometry for field along the [101], [010], and [001] directions are negative at all temperatures, and linear except for the [101] data [Fig.~S\ref{fig:FePHall}(a) and (c)] below 50~K. Fits to a two-band model show that the most significant drop in the hole-electron mobility ratio is in the 25-50~K range [Fig.~S\ref{fig:FePHall}(b)], i.e. around \textit{T*}, further proof of a change in the electronic properties of FeP at low temperatures.

\subsection{Electronic Structure and Quantum Oscillations}

\begin{figure}
    \centering
    \includegraphics[width=1.0\columnwidth]{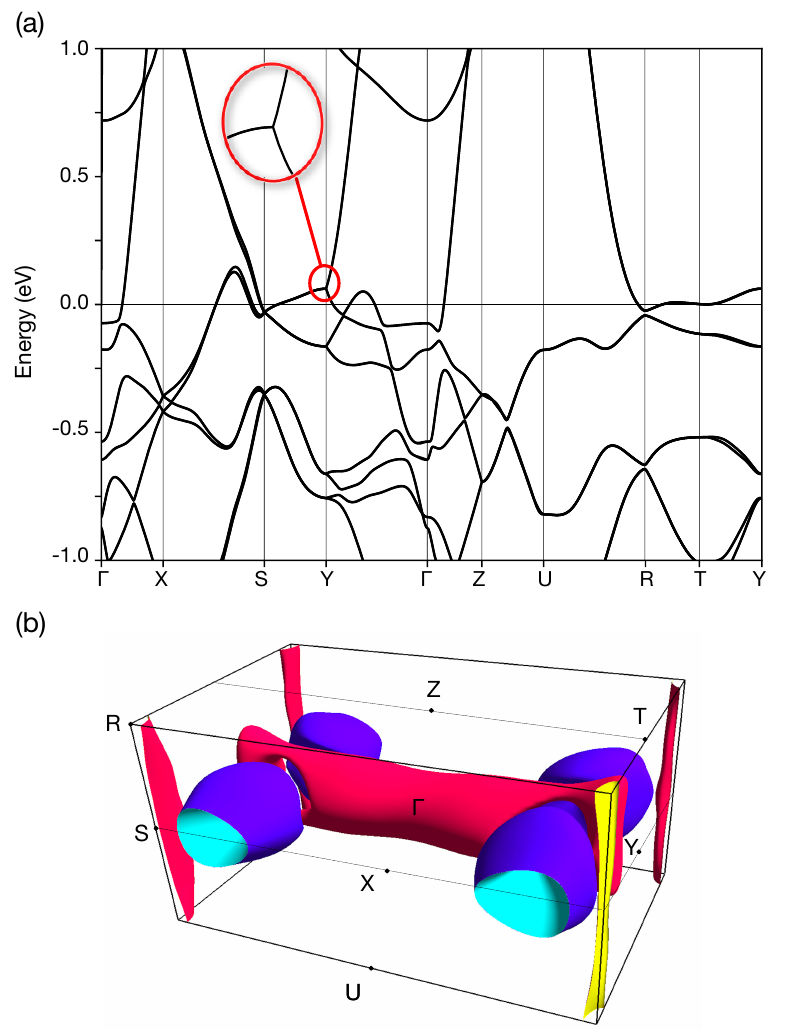}
    \caption{Density functional theory results for FeP. (a) Electronic structure of FeP calculated along high-symmetry lines labeled in panel (b). The zoom highlight shows the semi-Dirac point near 71~meV. (b) Calculated Fermi surface including all eight bands that cross $E_F$.}
    \label{fig:theory}
\end{figure}

To understand the role of FS geometry and complexity in the observed anomalous MR field and angle dependence, we present a comparison of QO data and calculated FS. Figure~\ref{fig:theory}(a) presents the electronic structure calculated along the high-symmetry lines in reciprocal space. We assumed a paramagnetic configuration because of the small magnetic moment seen in susceptibility \cite{WesterstrandhFeP}, which is a consequence of the octahedral crystal field that inverts the $4s$ and $3d$ energetic levels and changes the Fe configuration from commonly observed $d^{6}$ to $d^{8}$. As will be seen, a comparison to experiment shows good agreement with the paramagnetic Fermi surface at low temperature. The band structure is similar to that recently reported \cite{ChernyavskiiFeP}, though our Fermi level appears to be at a slightly lower energy, which somewhat alters the appearance of the Fermi surface [Fig.~\ref{fig:theory}(b)]. As expected from the angular dependence of MR, the shapes of the FS pockets are not simple. The nonsymmorphic symmetry of the MnP-type structure implies the presence of several linear bands and semi-Dirac points. While the eightfold-degenerate anisotropic Dirac points at R and S are split via the spin-orbit coupling in analogy to other topological pnictides \cite{kane, cuono_eur}, the fourfold-degenerate points at X, Y and Z are protected by the nonsymmorphic symmetry \cite{CuonoMnPType}.
 
The calculated Fermi surface is visualized in Fig.~\ref{fig:theory}(b). As expected from the angular dependence of magnetoresistance, there are several pockets (of both hole and electron nature) and they have a complicated geometry. As the nonsymmorphic symmetry induces non-trivial features that may strongly affect transport properties, it is particularly important to examine the agreement between the theoretical FS and experimental data. To this end, we have calculated the frequencies of quantum oscillations which arise from orbits or carriers perpendicular to an applied magnetic field around the perimeters of FS pockets.

\begin{figure*}
    \centering
    \includegraphics[width=.85\textwidth]{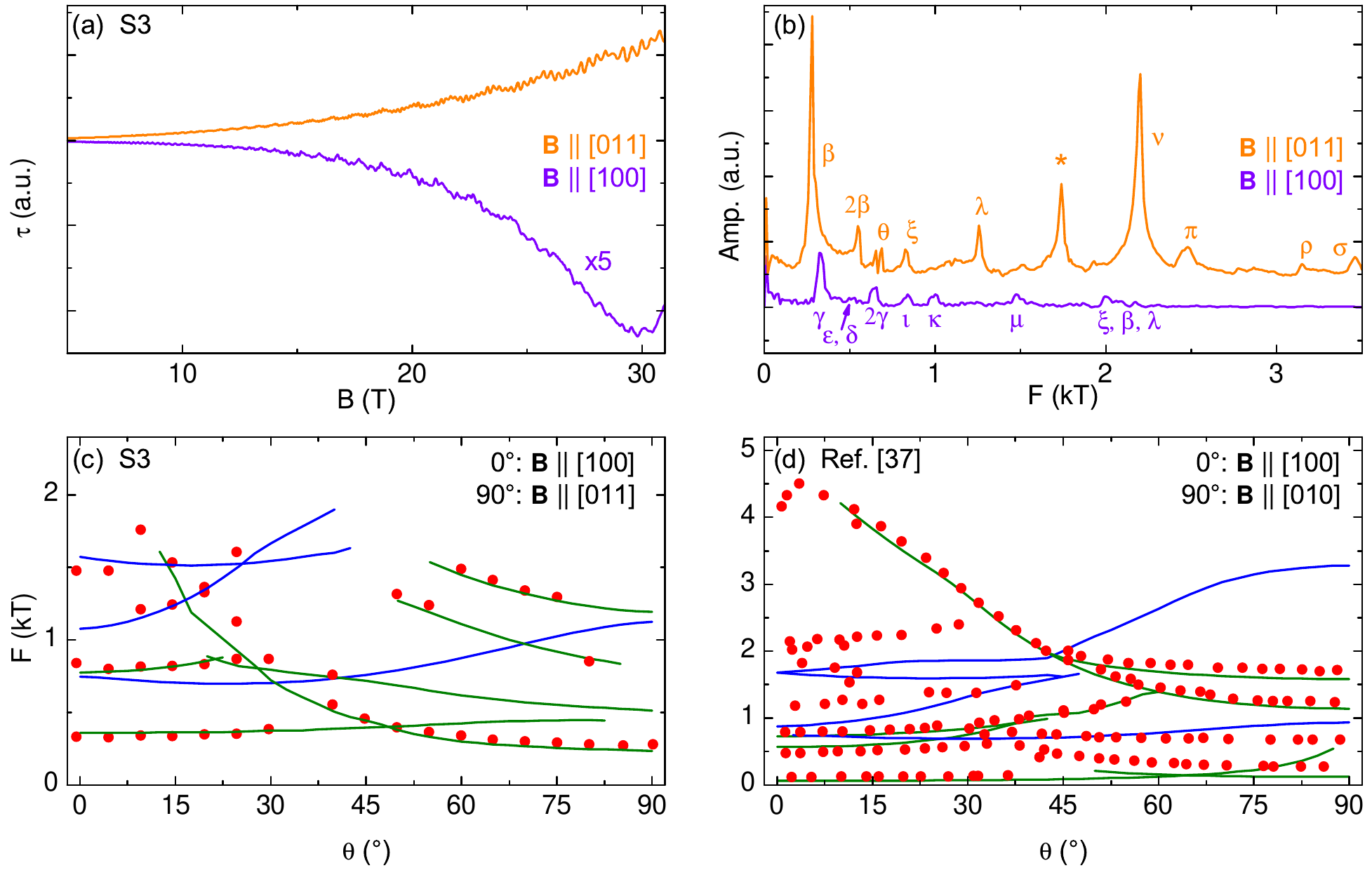}
    \caption{Experimental-theoretical comparison of quantum oscillations in FeP. (a) Magnetic torque data taken at high field on sample S3 for field applied along raw data for \textbf{B}~$\parallel$~ [011] (orange) and the \textit{a}-axis (purple, amplitude increase five times). (b) Quantum oscillation frequency spectra of the data in (a). Peaks marked with Greek letters correspond to those found at similar frequencies ($\pm$100~T) and the same angle in a prior quantum oscillation study \cite{NozueFePQOs}, using the notation of that work. A peak not identified in the previous study is labeled with an asterisk. (c) A comparison of observed fundamental quantum oscillation frequencies (red dots) to predicted electron (green lines) and hole (blue lines) band oscillation frequencies for $E_F = +0.15$~meV generated by \textsc{skeaf} calculations for rotation between the angles in (a) and (b) of S3. (d) A comparison of \textsc{skeaf} generated frequencies from the same FS to data from Nozue \textit{et al}. \cite{NozueFePQOs} for field rotated between the \textit{a}- and \textit{b}-axes.}
    \label{fig:QOs}
\end{figure*}

The extremal orbital frequency, with units of magnetic field, is related to the area enclosed by an orbit, thus the shape of the Fermi surface determines QO frequencies. While QOs are visible in the MR data, we found them easier to detect and analyze with torque magnetometry, which has a higher sensitivity in materials with low resistance and simpler background; our previous QO work on this family showed torque oscillations as low as 5~T when they did not appear in resistance up to 30~T \cite{CampbellFeAs, CampbellCoAs}. Magnetic torque was measured in parallel with MR measurements of sample S1, but on a different sample (S3). An example of the raw torque signal is shown in Fig.~\ref{fig:QOs}(a) for \textbf{B}~$\parallel$~\textit{a} and \textbf{B}~$\parallel$~[011]. Because the torque amplitude becomes very small when the field is aligned with crystal axes, the \textbf{B}~$\parallel$~\textit{a} data have been multiplied by a factor of five. Oscillations are clear down to about 7~T for both angles. A polynomial fit was subtracted to remove the non-oscillatory background, and a fast Fourier transform was performed on the residual signal. The frequency spectra for the same two angles are in Fig.~\ref{fig:QOs}(b). The Greek letters in that panel correspond to orbits identified in a previous QO paper that rotated between the three principal axes \cite{NozueFePQOs}, based on their having similar frequency values. Our experiment reproduces the same reported frequencies at the two common angles, but there is one (marked with an asterisk) not previously observed. The extreme similarity at both angles justifies using the previous results as a second verification of our band structure calculations. We also tracked the change in amplitude of the most prominent peaks with temperature in order to calculate the effective mass for field along [011] and [100]. Those data (with, where possible, a comparison to previous results and theoretical predictions) are in the SI [Table~II].

Figures ~\ref{fig:QOs}(c)-(d) present a comparison between theoretical and experimental frequencies identified by angular sweeps in this (c) and in the prior (d) study \cite{NozueFePQOs}. In the \textsc{skeaf} calculations, the Fermi level was tuned over a wide range of energies and the best agreement was found when $E_F$ was set to only 15~meV above the original density functional theory (DFT) value. The agreement in a 5~meV window around this value was noticeably better than at any alternative energy setting. 

We note that only the fundamental frequencies were included in the plot; as noticed both by us and in the prior work, there are many higher frequencies in FeP that can be attributed to harmonics or magnetic breakdown, and which \textsc{skeaf} would therefore not predict. Similarly, predicted frequencies that were unlikely to be observed, either because the predicted effective mass was too large, they were very close in frequency to another band, or they existed over a narrow angular range, have been excluded. In (c) some of the bands have been rigidly shifted by up to 200~T, but this does not change the qualitative angular dependence. It is possible in both cases to identify similarities in the angular dependence of the theoretical and experimental frequencies, confirming in multiple dimensions general agreement between the theoretical and experimental Fermi surfaces, with minimal adjustment needed. This supports our use of the paramagnetic Fermi surface as a reference point. The agreement seems overall to be much better with the predictions for electron (green lines) rather than hole (blue) bands.

%%%%%%%%%%%%%%%%%%%%%%%%%%%%%%%%%%%%%%%%%%%%
\section{\label{sec:Discussion}Discussion}

There are two distinct interesting aspects of the magnetotransport of FeP, 1) that it is very large and 2) that it follows a completely linear field dependence only for \textbf{B}~$\parallel{} c$-axis, and that it does so from very low field up to more than 30~T, an unprecedented range. These phenomena do not seem intertwined, since the MR is large and nonsaturating at all angles. Linear MR, meanwhile, occurs only at a specific orientation. However, studies of topological semimetals have shown that the two can have roots in the same physics. Electron-hole compensation generally leads to a large, nonsaturating MR \cite{PippardMRinMetals}, and can occur when the valence and conduction bands touch close to $E_F$. These small pockets near the Fermi level can often have a linear dispersion, i.e. the touching points are topological Weyl or Dirac points. Cd$_3$As$_2$ \cite{FengCd3As2} and TlBiSSe \cite{NovakTlBiSSe} are two materials that combine large MR, linear MR, and topological band structure features.

Large MR is seen in many more MnP-type materials than linear MR. As noted, the MR of FeP is comparable to observations in MnP \cite{TakaseMnPHiField} and CrP \cite{NiuCrP}. While smaller, CoP [SI, Fig.~S\ref{fig:CoPTransport}] still has a sizable increase in comparison to CoAs. None of these compounds show any sign of MR saturation. In the semiclassical picture of MR, resistivity should stop increasing with field only at very high \textit{B}, if at all, for compensated or nearly compensated materials where electron and hole transport is balanced \cite{PippardMRinMetals}. Analyzed with a two band model, Hall effect results for S1 show a two order of magnitude change in the electron-hole mobility ratio toward parity starting at 50~K [SI, Fig.~S\ref{fig:FePHall}], roughly in the region of \textit{T*}. The preservation of band structure features by the space group means other B31 pnictides will have a similar dispersion, providing the conditions for large MR in those that have low residual resistivities (generally, the phosphides).

There are not many cases of linear MR over such a wide range of field. Very few have as large of an increase, and of those none show linearity to as low of field. Recently, Ru$_2$Sn$_3$ was reported to have a linear MR starting from low field at certain orientations \cite{WuRu2Sn3}. However, there is still a visible initial quadratic component not seen in FeP, meaning there is a different process at work (the authors suspect a gap opening in the low carrier, semimetallic band structure). Bilayer graphene \cite{KisslingerGrapheneLMR} and silver chalcogenides \cite{HusmannLMRAgSe} can in fact exhibit such behavior beyond 60~T. But the weak temperature dependence and 2D nature of the graphene samples likewise point to a wholly different origin.  The high-$T_c$ cuprates are another example \cite{Giraldo-GalloCuprateLMR}, but the low field behavior is obscured by the superconducting state, so it is difficult the point to where normal state linear MR begins. The lack of linear $\rho{}$(T) in FeP also prohibits a quantum criticality explanation like that applied to the cuprates. Linear MR has been seen in transition metal pnictides \cite{XuAgChs,FengCd3As2}, even at pulsed field in Cd$_3$As$_2$ \cite{NarayananCd3As2}, but the explanation in that case is mobility fluctuations due to disorder. The very low residual resistivity of S1 rules this out, and we still see linearity (with a smaller MR) in a sample with much lower RRR [SI, Fig.~S\ref{fig:MoreHiField}]. The fact that linear MR is still present in that sample also shows that the MR field dependence is preserved against disorder, even when the magnitude is not. Weyl semimetal candidates PrAlSi and LaAlSi show quasilinear MR up to 9~T \cite{LyuPrAlSi}. Both show similar behavior, despite a difference in magnetism and the appearance of quantum oscillations. Analysis of the Hall data of PrAlSi shows that this can likely be attributed to compensated behavior and is not inherently topological.

The explanation given for linearity under pressure in CrAs (where MR roughly doubles up to 14~T) was based on the assumption that only a single, small FS pocket contributes to magnetoconductivity for \textbf{B}~$\parallel$~\textit{c}-axis. This is the Abrikosov quantum linear MR picture, which says such a pocket can very quickly be reduced to the lowest Landau level in field. The linear dispersion leads to a vanishing effective mass \cite{AbrikosovLinearMR}, producing a linear MR from low field. This theory cannot be applied to FeP, as our DFT calculations revealed the semi-Dirac point about 71~meV above the Fermi level [Fig.~\ref{fig:theory}(a)] while based on the QO comparison, the Fermi level in the analyzed samples seems to be shifted up by 15~meV. This means it is further away than the semi-Dirac point in CrP (found at -47~meV) which showed large but nonlinear MR \cite{NiuCrP,NiuCrAs}, whereas at optimal pressure the same point was less than 10~meV below $E_F$ in CrAs. %Furthermore, the Fermi velocities in FeP are too small (and so, the effective masses too large) for this theory to be applicable. %can we make the point on Fermi velocities more quantitative?

It does not seem possible to explain linear MR in FeP solely by comparison to previous experimental work. We look instead to a recent theoretical work which stated that in systems with both topological band structure features and long range magnetic order, a positive linear magnetoresistance can emerge from low field \cite{XiaoLMR} and be maintained in higher fields. This effect is the result of intra-band scattering of the topological band, and is predicted to be comparable to and potentially even larger than the inter-band scattering contribution. A large amount of this intra-band scattering would explain the linear MR in FeP, with more leeway for the location of the half-linear dispersion, and its absence in paramagnetic CrP \cite{NiuCrP}, CoAs \cite{CampbellCoAs}, and CoP [SI, Fig.~S\ref{fig:CoPTransport}]. The more stringent requirements on the magnetic state explain why linear MR is not as widespread as large MR in the B31 class, in spite of crystallographic protection of topological points. The relevance of linear MR only appearing for field along the helimagnetic propagation direction is unclear, though it could also be that this effect appears at more orientations, but is obscured by other, larger contributions. Relating back to the intrascattering idea, the exact angle between applied field, magnetic moments, and topological dispersions could affect the ratio of linear contributions to others.

We note that the MR is minimized for field along the \textit{c}-axis, and still tends toward high field linearity at other angles, a possible sign of dominance of the semi-Dirac point-driven MR at high field. Whether this explanation could also apply to CrAs is uncertain, as MR becomes closest to linear only after suppression of magnetic order \cite{NiuCrAs}. However, even at ambient pressure the MR in CrAs is approximately linear by 5~T. The low field behavior changes from sublinear to roughly quadratic with higher pressure. This could also be an example of changing weight of multiple MR contributions with different pressure dependencies. Linear MR has been seen in the other two magnetic B31 compounds in certain field ranges: FeAs shows a quadratic-linear crossover around 10~T \cite{CampbellFeAs}, while MnP can have quasilinear MR from below 2~T \cite{TakaseMnPHiField}. However, interpretation of the latter data is complicated by the presence of metamagnetic transitions, which may alter the relation between the semi-Dirac point and magnetic order.

%\section{\label{sec:Conclusion}Conclusion}

The magnetoresistance of FeP is large and nonsaturating at all angles, but with significant changes in magnitude and field dependence based on orientation. At most angles this large MR has a superlinear dependence that straightens out by 35~T. However, with field along the \textit{c}-axis, MR is linear from very low field. We believe that this can result from the combination of ordered magnetism and topological band structure present in FeP, which produces an anisotropic linear response. A key finding is that large and linear MR in FeP have separate origins. Other materials in this family display large (e.g. CrP) or linear (CrAs) MR, but not both; in combining the two, FeP exhibits impressive low temperature, high field behavior. The field dependence is large, robust to disorder, and simple (lacking even quantum oscillations at low temperature), which could be very useful for future application. Given what has already been achieved with CrAs and MnP, the behavior of the magnetic state and linear MR under applied pressure is an interesting path for followup work.

%%%%%%%%%%%%%%%%%%%%%%%%%%%%%%%%%%%%%%%%%%%%
\section{\label{sec:Methods}Methods}

\textbf{Crystal growth.} Samples were grown by the chemical vapor transport (CVT) method using I$_2$ as the transport agent \cite{BellavanceFePPreparation, RichardsonFePCVT, WesterstrandhFeP, BinnewiesCVTBook, NozueFePQOs}. 
For the CVT growth, a single zone horizontal tube furnace was used with the middle at about 900~$\degree{}$C, and the end about 200~$\degree{}$C cooler due to the natural gradient of the furnace. Fe and P powder along with I$_2$ polycrystals were sealed under vacuum in a tube half the length of the furnace, arranged such that the material was initially at the hot end of the tube (the furnace center). Single crystals of FeP were found at the cold end of the furnace after 10-14 days, often with FeP powder; on occasion some would also be found at the hot end. The crystals grown with CVT were polyhedral with dimensions of roughly 0.5-3~mm in each direction, but nevertheless showed a clear preference for growing along the \textit{b}-axis, the shortest crystallographic axis. X-ray diffraction (XRD) of ground single crystals showed single phase FeP with lattice parameters of \textit{a}~=~5.10~$\text{\AA{}}$, \textit{b}~=~3.10~$\text{\AA{}}$, and \textit{c}~=~5.79~$\text{\AA{}}$, in line with previous reports \cite{SelteFeP, NozueFePQOs}. We also tried growing crystals with prereacted FeP powder and a similar technique, but the samples seemed to be of lower quality than those grown from the raw elements, as they had lower RRR and MR values.

\textbf{Physical property measurements.} Single crystals were aligned with a combination of single crystal x-ray diffraction and Laue photography. In-house electrical transport measurements up to 14~T were taken in a Quantum Design Physical Properties Measurement System. Those to 31 and 35~T were made using two different resistive magnets at the National High Magnetic Field Laboratory (NHMFL) and a rotating sample platform, in conjunction with torque measurement done via piezoresistive cantilever.

\textbf{Density functional theory calculations.} First-principles calculations based on DFT were performed using \textsc{Quantum Espresso} package \cite{qe1,qe2}. We treated the exchange and correlation interaction within the generalized gradient approximation (GGA) \cite{pbe}, and the ion-electron interaction with the projector augmented-wave pseudopotentials from the pslibrary database \cite{kresse-joubert, pslibrary}. The electron wave functions were expanded in a plane wave basis set with the cutoff of 50 Ry. The FeP structure was modeled by the orthorhombic unit cell with lattice constants set to those obtained via x-ray. The internal degrees of freedom were relaxed until the forces on each atom became smaller than $10^{-3}$ Ry/bohr. The Brillouin zone sampling at the level of DFT was performed following the Monkhorst-Pack scheme using a $12\times16\times10$ k-points grid. Spin-orbit coupling (SOC) was taken into account self-consistently. The Fermi surfaces were calculated on the interpolated mesh of $60\times80\times50$ using the \textsc{Paoflow} code and visualized with FermiSurfer \cite{paoflow, fermisurfer}. The quantum oscillation frequencies expected for the calculated Fermi surface were evaluated using the Supercell K-space Extremal Area Finder (SKEAF) code \cite{JulianSKEAF}.

\section{\label{sec:Contributions}Data Availability}
All relevant data are available from the authors upon reasonable request.

\section{\label{sec:Acknowledge}Acknowledgments}

This work was supported by the National Science Foundation Division of Materials Research award no.~DMR-1905891 and the Gordon and Betty Moore Foundation's EPiQS Initiative through grant no.~GBMF9071. A portion of this work was performed at the National High Magnetic Field Laboratory, which is supported by National Science Foundation Cooperative Agreement nos.~DMR-1157490 and 1644779 as well as the state of Florida. D.J.C. acknowledges the support of the Anne G. Wylie Dissertation Fellowship. We also acknowledge the support of the Maryland NanoCenter and its FabLab. C.A. is supported by the Foundation for Polish Science through the International Research Agendas program co-financed by the European Union within the Smart Growth Operational Programme. The computational resources at the Texas Advanced Computing Center of the University of Texas, Austin are gratefully acknowledged.

\section{\label{sec:Contributions}Author Contributions}
D.J.C. and J.C. grew the samples and conducted zero and low magnetic field measurements. D.J.C., K.W., B.W., Y.S.E., P.N., and D.G. made the high field measurements. J.S., C.A., and L.W. performed theoretical calculations. M.B.N. and J.P. supervised the project. D.J.C., J.S., and J.P. wrote the manuscript with input from all authors.

\section{\label{sec:Contributions}Competing Interests}
The authors declare no competing interests.

\bibliographystyle{naturemag}
\bibliography{FePRefs}

\end{document}